\documentclass[twocolumn,showpacs,showkeys,prd]{revtex4-1}

\usepackage{array}
\usepackage{amsmath,amsfonts,amssymb,dsfont,mathrsfs,centernot}
\usepackage{graphicx}
\usepackage{subfigure}
\usepackage[colorlinks=True,linkcolor=blue,citecolor=blue]{hyperref}

\usepackage{amsmath,amssymb,amsfonts,dsfont,mathrsfs,amsthm}
\usepackage{centernot}
\usepackage{hyperref}
\usepackage{xcolor}
\usepackage{comment}
\hypersetup{linktocpage,colorlinks=true,urlcolor=green,linkcolor=blue,citecolor=red}
\usepackage{units}
\usepackage{array}

\newcommand{\abs}[1]{\left|#1\right|}

\newcommand{\cdf}{{\boldsymbol{\mathcal{D}}}}
\newcommand{\covd}{\mathcal{D}}

\newcommand{\df}{{\bf{d}}}

\newcommand{\fy}{\centernot}

\newcommand{\ga}{\gamma}

\newcommand{\Lag}{\mathscr{L}}

\newcommand{\Or}{\mathscr{O}}

\newcommand{\Ri}{\mathcal{R}}

\newcommand{\w}{\wedge}

\newcommand{\bps}{\ensuremath{\bar{\psi}}}
\newcommand{\Bps}{\ensuremath{\bar{\Psi}}}

\newcommand\VIF[1]{\hat{\bf{e}}^{\hat{#1}}}

\newcommand\hvif[1]{\hat{\bf{e}}^{{#1}}}

\newcommand\SPIF[2]{\hat{\boldsymbol{\omega}}^{\hat{#1}}{}_{\hat{#2}}}

\newcommand{\RIF}[2]{\hat{\bf{R}}^{\hat{#1}}{}_{\hat{#2}}}
\newcommand{\hRif}[2]{\hat{\bf{R}}^{{#1}}{}_{{#2}}}

\newcommand{\TF}[1]{\hat{\boldsymbol{\mathcal{T}}}^{\hat{#1}}}

\newcommand{\hcont}[3]{\hat{\mathcal{K}}_{#1}{}^{#2}{}_{#3}}

\newcommand{\CONTF}[2]{\hat{\boldsymbol{\mathcal{K}}}^{\hat{#1}}{}_{\hat{#2}}}

\newcommand{\beq}{\begin{equation}}
\newcommand{\eeq}{\end{equation}}
\newcommand{\ber}{\begin{eqnarray}}
\newcommand{\eer}{\end{eqnarray}}

\renewcommand{\(}{\left(}
\renewcommand{\)}{\right)}
\renewcommand{\[}{\left[}
\renewcommand{\]}{\right]}

\newcommand{\dn}[2]{\,{\rm{d}}^{#1}{#2}\;}

\hypersetup{pdftitle={Phenomenological Constraints to Dimensionality of the Spacetime with Torsion},pdfauthor={Oscar Castillo-Felisola},pdfkeywords={Extra Dimensions, Modified Gravity},pdflang={English}}

\begin{document}

\title{Phenomenological Constraints to Dimensionality of the Spacetime with Torsion}

\author{Oscar Castillo-Felisola}
\email{o.castillo.felisola@gmail.com}

\author{Crist\'obal Corral}
\email{cristobal.corral@postgrado.usm.cl}

\author{Iv\'an Schmidt}
\email{ivan.schmidt@usm.cl}

\author{Alfonso Zerwekh}
\email{alfonso.zerwekh@usm.cl}

\affiliation{Departamento de F\'\i sica y Centro Cient\'\i fico Tecnol\'ogico de Valpara\'\i so, Universidad T\'ecnica Federico Santa Mar\'\i a, Casilla 110-V, Valpara\'\i so, Chile.}

\begin{abstract}
  Using the recent limits established by ATLAS to the contact four-fermion interaction, bounds on the size of the extra dimensions of space-time have been found, by assuming that the contact interactions come through the inclusion of torsion in the higher dimensional theory.   For  two extra dimensions, the limits are comparable to those in the literature, while for higher dimensional space-time, the volume of the extra dimensions is strongly constrained.
\end{abstract}

\pacs{04.50.-h,04.62.+v,11.25.Mj,11.25.-w}
\keywords{Extra dimensions, Torsion, Modified Gravity}

\maketitle

\section{Introduction}\label{sec:intro}

The Standard Model (SM) of particle physics has proved to be a very successful theory, but still has problems. Among these, one that is considered fundamental is the {\it Hierarchy Problem}, which might be solved by considering extra dimensions. In higher dimensional space-times the gravitational theory of Einstein can be generalized in several ways. The simplest of these generalizations is due to Cartan.

Cartan's generalization of general relativity, which introduces torsion to the picture, can be coupled to fermionic matter in a natural way. Since the equation of motion for the spin connection is a constraint, related to the contortion, it can be used to get rid of the torsion in the original action. The new action contains standard general relativity and matter fields with an additional contact four fermion interaction\cite{SUGRA-book}.  

The effective four fermion interaction term has a coupling constant proportional to Newton's gravitational constant, $G_N\sim M_{pl}^{-2}$. Therefore, at first glimpse this interaction is highly suppressed. Nevertheless, in the last twenty years diverse scenarios have proposed that the existence of extra dimensions could explain the hierarchy problem, and thus the (higher dimensional) fundamental gravity scale might be roughly $M_*\sim \Or(1) \;\unit{TeV}$\cite{ADD1,ADD2,RS1,RS2}. Recently, limits to the size of the extra dimensions have been set up  by direct searches of 
quantum black holes\cite{Gingrich:2012vb} and the influence of the exchange of virtual gravitons on dilepton events\cite{:2012cb}.

On the other hand, the ATLAS collaboration has presented experimental limits for the coupling constant of four fermion contact interaction\cite{Aad:2011aj,ATLAS:2012pu,:2012cb}. This result is particularly important for imposing bounds on the value of the fundamental gravity scale, $M_*$, and by extension in order to find limits on the dimensionality or size of the space-time.

The aim of this work is to find natural bounds and limits on the typical size of the extra dimensions, as well as on the gravitational scale, $M_*$. In Sec. \ref{sec:CEF}, for the sake of completeness,  a brief presentation of Cartan's generalisation of gravity coupled with fermions is presented. In the next sections constraints are found for different higher dimensional scenarios, such as those proposed by Arkani-Hamed, Dimopoulos and Dvali\cite{ADD1,ADD2} in Sec. \ref{sec:ADD}, and by Randall and Sundrum\cite{RS1,RS2} in Sec. \ref{sec:RS}.  In Sec. \ref{sec:res} a summary of results and conclusions are presented.

\section{Cart\'an-Einstein Gravity with Fermions}\label{sec:CEF}

It is a well-known fact that Einstein's gravity is a field theory for the metric, and the space-time connection is required to be torsion-free. 

However, when the first order formalism of pure gravity is considered, whose fields are the vielbeins and spin connection, the torsion-free imposition is nothing but the equation of motion for the spin connection.

Moreover, for gravity coupled with matter (specially fermionic matter), this condition changes and introduces a four-fermion contact interaction. The modification of the fermionic Lagrangian due to the presence of torsion is presented below.

The Cartan-Einstein action in $D$-dimensions is,
\begin{align}
  S_{gr} = \frac{1}{2\kappa^2}\int\frac{\epsilon_{\hat{a}_1\cdots \hat{a}_D}}{(D-2)!}\hRif{\hat{a}_1 \hat{a}_2}{}\w\hvif{\hat{a}_3}\w\cdots\w\hvif{\hat{a}_D},\label{CE-action}
\end{align}
where $\VIF{a}$ and $\SPIF{a}{b}$ are the vielbein and spin connection 1-forms respectively, and $\TF{a}$ and $\RIF{a}{b}$ are the torsion and curvature 2-forms satisfy the structure equations
\begin{align}
  \df\VIF{a}+\SPIF{a}{b}\w\VIF{b} &= \TF{a},\label{struc.eq.1}\\
  \df\SPIF{a}{b}+\SPIF{a}{c}\w\SPIF{c}{b}&=\RIF{a}{b}.\label{struc.eq.2}
\end{align}
Additionally, the Dirac action in arbitrary dimension is
\begin{align}
  S_\Psi ={}& -\int \frac{\epsilon_{\hat{a}_1\cdots \hat{a}_D}}{(D-1)!} \Bps \hvif{\hat{a}_1}\w\cdots\w\hvif{\hat{a}_{D-1}}\ga^{\hat{a}_D}\hat{\cdf}\Psi \notag\\
 & -m\int\frac{\epsilon_{\hat{a}_1\cdots \hat{a}_D}}{D!}\Bps \hvif{\hat{a}_1}\w\cdots\w\hvif{\hat{a}_{D}}\Psi,
\end{align}
with $\hat{\cdf}$ the exterior derivative twisted by the spin connection.

Since the total action is 
\begin{align}
  S = S_{gr} + S_{\Psi},
\end{align}
the equations of motion for the whole system are,
\begin{align}
  \hat{\Ri}^{\hat{m}}{}_{\hat{a}_3} -\frac{1}{2}\hat{\Ri} \delta^{\hat{m}}_{\hat{a}_3} &=\kappa^2\bar{\Psi}\[\ga^{\hat{m}}\hat{\covd}_{\hat{a}_3}-\delta^{\hat{m}}_{\hat{a}_3}\(\fy{\hat{\covd}}+m\)\]\Psi, \label{ED-eom}\\
  \hcont{\hat{a} \hat{b} \hat{c} }{}{} &= -\frac{\kappa^2}{4} \bar{\Psi}\ga_{\hat{a} \hat{b} \hat{c}}\Psi. \label{cont-constraint}
\end{align}
Since Eq. \eqref{cont-constraint} is a constraint, it can be substituted into the action, using that
\begin{align}
  \SPIF{a}{b} \mapsto \SPIF{a}{b}+\CONTF{a}{b},
\end{align}
where the former is a general spin connection, while the later is a torsion-free one.

The new action is
\begin{align}
  S ={}& \int\dn{D}{x}\abs{\hat{e}}\[\frac{1}{2\kappa^2}\hat{\Ri} -\Bps\(\fy{\hat{\covd}}+m\)\Psi\right.\notag\\
  & \left.+\frac{\kappa^2}{32}\bar{\Psi}\ga_{\hat{a} \hat{b} \hat{c}}\Psi\bar{\Psi}\ga^{\hat{a} \hat{b} \hat{c}}\Psi\],
\end{align}
which is a torsion-free theory of gravity coupled to a fermion with a four-fermion contact interaction.

In the next sections the gravitational aspect of this model is not considered, based in the fact that the universe is essentially flat, and gravitational forces are extremely weak in comparison to other known interactions. Nonetheless, the existence of torsion still leaves behind a four-fermion interaction. Hereon, our objective will be to compare the four-fermion interaction coming from this gravity model with the four fermion interaction limits found in ATLAS experiment.

\section{Bounds on Four-Fermion Interaction}\label{sec:4fi}

Early proposals of contact four-fermion interaction signals in colliders are found in \cite{RevModPhys.56.579,RevModPhys.58.1065,Chiappetta:1990jd}. These works inspired searches at the Large Hadron Collider (LHC) experiments. In particular, the ATLAS  collaboration has found limits on the scale of four-fermion contact interaction, by analyzing the invariant mass and angular distribution of dijets\cite{Aad:2011aj,ATLAS:2012pu,:2012cb}. The strongest constraint comes from an interaction of chiral fermions in the form
\begin{align}
  \Lag_{qqqq} = \pm\frac{g^2}{2\Lambda^2}\; \bps_{qL}\ga^a\psi_{qL}\bps_{qL}\ga_a\psi_{qL},\label{Lag-qqqq}
\end{align}
which is experimentally excluded for $\Lambda<\unit[9.5]{TeV}$, assuming that the coupling constant $g\sim\Or(1)$. 

The aim of this section is to compare the above result  with the  four-fermion interaction coming from the Cartan-Einstein gravity action\footnote{Note that four-dimensional fermions are denoted with lower case symbol $\psi$, whilst the fermion in arbitrary dimension is denoted with the capitalized one, $\Psi$.}, 
\begin{align*}
  \Lag_{4\Psi} = \frac{\kappa^2}{32}\; \Bps\ga_{\hat{a} \hat{b} \hat{c}}\Psi \Bps\ga^{\hat{a} \hat{b} \hat{c}}\Psi.
\end{align*}
It becomes clear that if one starts in four-dimensions, where $\kappa^2\sim M_{pl}^{-2}$, $\ga_{abc}\sim \ga_d\ga^*$ and $\Psi = \psi_L +\psi_R$,  the comparison implies $\Lambda\sim M_{pl}\ggg \unit[10]{TeV}\;$, which is tautological.

Nonetheless, there exist models where the fundamental scale of gravity is not $M_{pl}\sim 10^{18}\;\unit{GeV}$, but rather a much lower one, $M_*$, which could be of order of the electro-weak scale, i.e. $M_{*}\sim M_{EW}$, giving a natural solution to the {\it Hierarchy Problem}. These models, however,  require  extra dimensions. Therefore, in the following the space-time will be considered to be a $(4+n)$-dimensional, where the $n$ extra dimensions are either compact or extended, and differentiation between models is made according to this characteristic. 

In order to achieve the goal of comparison, it is necessary to reduce the dimension of the space-time from $D$ down to four. Although the dimensional reduction could be a complex procedure\footnote{The difficulty for finding the effective theory comes from the fact that in higher dimensional space-times, the spinorial representation of the Lorentz group could have dimension different than four. Therefore, one should be aware of the decomposition of the spinors (including the profiles through the extra dimensions), as well as the Clifford algebra elements, prior to the integration of the extra dimensions.}, we will sketch how a term like Eq. \eqref{Lag-qqqq} appears.

First, in the contraction $\ga_{\hat{a} \hat{b} \hat{c}}\ga^{\hat{a} \hat{b} \hat{c}}$ one considers only the four-dimensional part, i.e., $\ga_{abc}\ga^{abc}$. Second, the antisymmetric product of three elements of the four-dimensional Clifford algebra is equal to the product of the missing element of the Clifford algebra times the chiral element, i.e., $\ga_{abc} \sim \ga_d \ga^*$. Next, the higher dimensional spinor $\Psi$ can be decomposed as the product of the four-dimensional times $n$-dimensional spinors, 
\begin{align}
  \Psi(x,\xi) = \sum_i \(\psi^{(i)}_L(x) +\psi^{(i)}_R(x)\)\otimes \lambda_i(\xi).
\end{align}
Therefore, after integration of the extra dimensions, the effective four-dimensional theory would have a term of the desired form
\begin{align}
  \Lag_{\text{eff}} = \frac{\kappa^2_{\text{eff}}}{32}\; \bps_{qL}\ga^a\psi_{qL}\bps_{qL}\ga_a\psi_{qL}.\label{Lag-eff}  
\end{align}  
Dimensional  analysis gives two possibilities. Either the effective coupling constant is directly related to the fundamental gravitational scale, $\kappa_{\text{eff}}\sim M_*^{-1}$, or is related to the effective four-dimensional one $\kappa_{\text{eff}}\sim \(M'_{pl}\)^{-1}$, where $M'_{pl}$ is the redefined Planck mass after the dimensional reduction.

\subsection{ADD Models}\label{sec:ADD}

ADD models\cite{ADD1,ADD2}  consist of a four-dimensional space-time with a set of $n$  compact extra dimensions, with typical length $R$. Matter is confined to the four-dimensional space-time for energies below $\Lambda\sim \frac{1}{R}$, while gravity propagates through the whole space-time. This configuration allows to solve the hierarchy problem, because the natural scale for gravity is not the effective four-dimensional one, but rather the $(4+n)$-dimensional.

The relation between the fundamental gravitational scale, $M_*$, and the four-dimensional effective one, $M_{pl}$ is given by
\begin{align}
  M_{pl}^2 \sim M_*^{2+n} R^n.\label{rel-ADD}
\end{align}
Additionally, the coupling constant of the Einstein action is  $\kappa^2\sim \frac{1}{M_*^{2+n}}$. It is worthwhile to mention that ADD scenarios are restricted to $n\geq 2$, because of gravitational constraints\cite{ADD1}. 

From Eq. \eqref{rel-ADD}, it follows that the typical radii of the extra dimensions  are
\begin{align}
  R\sim 10^{\frac{30}{n}-17}\(\frac{\unit[1]{TeV}}{M_*}\)^{\frac{2}{n}+1}\;\unit{cm}.
\end{align}
Assuming that the scale of the four-fermion interaction, $\Lambda$, is essentially the fundamental scale of gravity, $M_*$, one finds sizes of the extra dimensions from roughly a few micrometers down  to  a few tens femtometers, as shown in the second column of Table \ref{Tab:ADD-R}. These results are a refinement of the original ADD claim.
\begin{table}
  \caption{Typical radius of the extra dimensions in ADD models. $R_{\text{max}}$ for $\Lambda\sim M_*$. $R_{\text{min}}$ for $\kappa^2_{\text{eff}}\sim \Lambda^{-2}$, assuming $M_*=\unit[100]{GeV}$.} \label{Tab:ADD-R}
  \begin{center}
    \begin{tabular}{>{$}c<{$}|>{$}c<{$}|>{$}c<{$}}
      n & R_{\text{max}}[\unit{m}]& R_{\text{min}}[\unit{m}]\\
      \hline
      2& 10^{-6} &  10^{-16}\\
      3& 10^{-11}&  10^{-17}\\
      4& 10^{-13}&  10^{-17}\\
      5& 10^{-15}&  10^{-17}\\
      6& 10^{-16}&  10^{-18}\\
      7& 10^{-16}&  10^{-18}
    \end{tabular}
  \end{center}
\end{table}

On the other hand, the effective coupling constant in four-dimensions, $\kappa_{\text{eff}}^2$, should be related directly with the fundamental scale of gravity, i.e., $\kappa^2_{\text{eff}}\sim M_*^{-2}$, or in a similar way as before
\begin{align}
  R\sim \(\frac{\Lambda}{M_*}\)^{\frac{2}{n}}\frac{1}{M_*}.
\end{align}
Hence, one finds another limit for the typical size of the extra dimensions, as shown on the third column of Table \ref{Tab:ADD-R}.

 
\subsection{Randall-Sundrum Models}\label{sec:RS}

When considering Randall-Sundrum scenarios of brane-worlds\cite{RS1,RS2}, with metric
\begin{align}
  \hat{g}_{\hat{\mu}\hat{\nu}} = e^{-2k r_c\abs{\xi}}\eta_{\mu\nu}\delta^\mu_{\hat{\mu}}\delta^\nu_{\hat{\nu}} + r_c^2 d\xi_{\hat{\mu}}d\xi_{\hat{\nu}},
\end{align}
the relation between the four-dimensional Planck mass, $M_{pl}$, and the fundamental (five-dimensional) Planck scale, $M_*$, is given by
\begin{align}
  M_{pl}^2 = \frac{M_*^3}{k}.
\end{align}
Since  a well-known modulus stabilization method would ensure that the product $k r_c\sim 10$\cite{Goldberger:1999uk}, the relation between the gravitational scales and the length of the extra dimension is found to be
\begin{align}
  M_{pl}^2 \sim \frac{M_*^3 r_c}{10}.\label{rel-RS}
\end{align}

Analyzing both limits as in the previous section, the limit on the extra dimension size is
\begin{align}
  \unit[10^{-13}]{m}< r_c <\unit[10^{19}]{m}.
\end{align}
The range is particularly  wide because there is a single extra dimension. Although brane-worlds of codimension higher than one have been considered\cite{Carroll:2003db,Parameswaran:2006db,Burgess:2008ka,Bayntun:2009im,Nierop-PhD}, without a carefully thought-out moduli stabilization process the bounds on the extra dimensions sizes are equal to those found in Sec. \ref{sec:ADD} (shown in Table \ref{Tab:ADD-R}).


\section{Concluding Remarks}\label{sec:res}

In this work we have used the limits on four-fermion chiral contact interactions obtained by the ATLAS collaboration in order to constrain the typical
size of eventual extra dimensions in models where gravity admits torsion in the bulk. For a codimension 2 we have found an upper bound for the radius of the extra dimensions of the order of $\unit[10^{-6}]{m}$, which is comparable to the limits obtained from direct search of the Kaluza-Klein excitations of the graviton \cite{PDG}. Nevertheless, for higher numbers of extra dimensions we find that the constrains are much more stringent. This is due to the fact that the fundamental gravitational scale, $M_*$, is related with the effective one, say $M'_{pl}$, through higher powers, reducing the dependence of the model on the size of the extra dimensions. This result provides an example of the possibility of testing non-trivial extensions of General Relativity using collider data.

\section*{Acknowledgements}

We would like to thank to C. Dib, N. Neills, J.C. Helo and A. Carcamo for fruitful discussions and encourage through the realisation of this work. 

This work was supported in part by Fondecyt Project No. 11000287, Fondecyt Project No. 1120346.

\appendix


\end{document}